\documentclass[preprint,amsmath,amssymb]{revtex4}
\usepackage{graphicx}
\begin{document}

\preprint{}

\title{Travelling Fronts, Pulses, and Pulse Trains in a 1D discrete Reaction-Diffusion System}
\author{Priyadarshi Majumdar}
\affiliation{Dept of Physics, Vidyasagar Evening College, Kolkata 700 006, INDIA}
 
\author{Avijit Lahiri}
 \email{a_l@vsnl.com}
\affiliation{Dept of Physics, Vidyasagar Evening College, Kolkata 700 006, INDIA}
\begin{abstract}
\noindent We follow up an earlier work (briefly reviewed below) to investigate the temporal stability of an exact travelling front solution, constructed in the form of an integral expression, for a one-dimensional discrete Nagumo-like model without recovery. Since the model is a piecewise linear one with an on-site reaction function involving a Heaviside step function, a straightforward linearisation around the front solution presents problems, and we follow an alternative approach in estimating a `stability multiplier' by looking at the variational problem as a succession of linear evolution of the perturbations, punctuated with `kicks' of small but finite duration. The perturbations get damped during the linear evolution, while the kicks amplify only the perturbations located at specific sites (the `significant perturbations', see below) with reference to the propagating front. Comparison is made with results of numerical integration of the reaction-diffusion system whereby it appears likely that the travelling front is temporally stable for all parameter values characterising the model for which it exists. We modify the system by introducing a slow variation of a relevant recovery parameter and perform a leading order singular perturbation analysis to construct a pulse solution in the resulting model. In addition, we obtain (in the leading order) a 1-parameter family of periodic pulse trains for the system, modelling re-entrant pulses in a one-dimensional ring of excitable cells.
\end{abstract}
\maketitle

\section{Introduction}

\begin{figure}[hb]
\centering
\includegraphics[height=6cm,width=9cm]{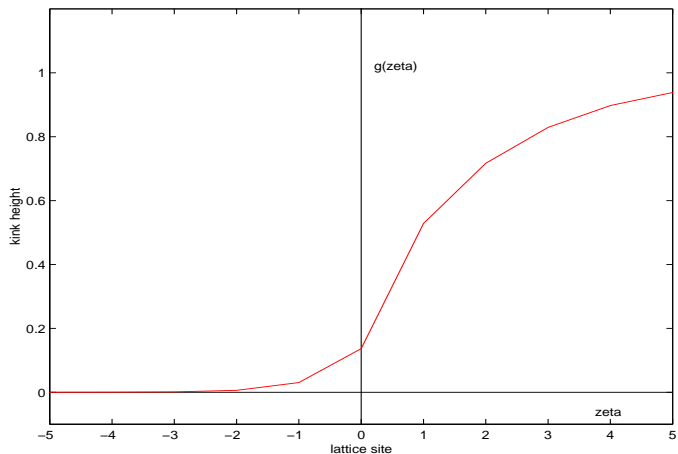}
\caption{\label{cap:f1}Kink profile function $g(\zeta)$ showing discontinuity of derivative at integer values of the propagation variable $\zeta$ (corresponding to integers representing lattice sites at $t=0$); parameter values are $w=0, D=1, a=0.1382$.}
\end{figure}

\noindent 
The discrete 
Nagumo (FN) 
model~\cite{ref1} without recovery on a $1D$ lattice
 reads
 \begin{subequations}
\begin{equation}
\frac{d u_{n}}{dt}=D(u_{n+1}
-2{u_n}+u_{n-1})+f(u_n),\label{eq:onea}
\end{equation}
\noindent where $D$ is a 
diffusion constant coupling
 adjacent lattice sites and
 $f$ is usually taken as a
 cubic bi-stable reaction
 function. It is considered relevant in numerous situations of interest, including the propagation of action potential along myelinated nerve axons, and excitations of cardiac cells(see~\cite{bell1,bell2}, and references in~\cite{ref1}). Following an 
earlier work~\cite{ref2} we
 consider a piecewise 
linear (PWL) version of 
\eqref{eq:onea}, with the reaction
 function given by 
\begin{equation} f(u) = - u - w 
+ \ominus (u-a),\label{eq:oneb} 
\end{equation}\end{subequations}\noindent where
 $\ominus$ stands for the 
Heaviside step function, the
 parameter $a~(<1/2)$ is an
 effective threshold 
characterising each site,
 separating the equilibrium
 values $0$ and $1$ of the
 order parameter $u_{n}$,
 and $w$ is a `recovery' variable
 (see below). The model \eqref{eq:onea}, \eqref{eq:oneb} is 
 essentially similar to the cubic 
Nagumo model in that each site is 
bi-stable in absence of the
 inter-site coupling. 
\noindent In~\cite{ref2} we gave
 an explicit construction of a
 travelling kink (or `front') 
solution to this PWL version
 of the FN equations, of the
 form
\begin{subequations}
\begin{equation}  
 u_n(t)=g(\zeta),\label{eq:twoa}
\end{equation}
\noindent where $\zeta $ is a
 propagation variable defined
 as $\zeta=(\chi t+n)$, $\chi$
 being the speed of the 
propagating front. The profile
 function $g$ was explicitly
 obtained as an integral 
expression 
\begin{equation}
g(\zeta)=a_p+\int_{0}^{\pi}{[b(\theta)e^{ip\theta}+b^*(\theta)e^{-ip\theta}]e^{-\lambda(\theta)(\zeta-p)}d\theta},\label{eq:twob}
\end{equation}
\end{subequations}
 where 
$p=[\zeta]$, the integer part
 of $\zeta$, and
 \begin{subequations}
 \begin{eqnarray}a_p=1-w-\frac{\gamma}{1+\gamma}\gamma^p~~(p\geq 0),\label{eq:threea}\\=-w+\frac{1}{1+\gamma}\gamma^{-p}~~(p\leq 0),\label{eq:threeb}\end{eqnarray}\end{subequations}\begin{eqnarray}b(\theta)=-\frac{1}{2\pi}\frac{1}{2D+1}\frac{1}{1-\nu cos\theta}\frac{1}{1-e^{-i\theta}e^{-\mu(1-\nu cos\theta)}},
\label{eq:four}\\ \lambda(\theta)= \frac{1}{\chi}(2D cos\theta-(2D+1)).\label{eq:five}
\end{eqnarray} 
\noindent Here the parameters $\gamma$, $\mu$, $\nu$ are defined as \begin{subequations}\begin{eqnarray}\gamma\equiv 1+\frac{1}{2D}-\sqrt(\frac{1}{D}+\frac{1}{4D^2}),\label{eq:gamma}\\\mu\equiv\frac{2D+1}{\chi},\label{eq:chi}\\\nu\equiv\frac{2D}{2D+1}.\label{eq:nu}\end{eqnarray}\end{subequations}\noindent The speed $\chi$ gets determined through a matching condition, for which we refer to~\cite{ref2}. An important observation is that, in contrast to a continuously distributed excitable medium, a travelling front on a discrete lattice gets {\it pinned} (in this context, see~\cite{ref1,Keener}) as the threshold approaches a certain limiting value $\tilde{a}$ (see~\cite{ref2}) where
\begin{eqnarray}
\tilde{a}=-w+\frac{1}{2}[1-\sqrt(\frac{1}{1+4D})]<\frac{1}{2}-w.\label{eq:a_tilde}
\end{eqnarray}
\noindent For a 1D continuously distributed reaction-diffusion system with the same reaction funcion as in eq. \eqref{eq:oneb} a  travelling kink solution would exist, for any given $D$, for all values of the threshold parameter in the range $0<a<\frac{1}{2}-w$.
\vskip .5cm 
\noindent The above explicit solution was
 obtained in~\cite{ref2} by noting that, as
 the front propagates, there
 occur successive time
 intervals (equivalently,
 intervals in $\zeta$) during
 which the system \eqref{eq:onea}, \eqref{eq:oneb}
 evolves linearly, and the
 transition from one interval
 to the next is marked by the
 value of $u_n$ crossing the
 threshold for some 
appropriate site $n$. The full
 solution is then obtained by 
an appropriate matching at
 these transition points between 
superpositions of the 
eigenmodes of the linear 
system.
\vskip .5cm
\noindent Fig. \ref{cap:f1} shows the pulse profile $g(\zeta)$ for arbitrarily chosen values of $a$, $D$, where one observes that $g(\zeta)$ is piecewise continuous, with a discontinuous derivative at each lattice site owing to the presence of the Heaviside step function in the model (we choose the origin of time such that $u_0$ crosses the threshold $a$ at $t=0$ and, accordingly, $u_n$ crosses the threshold at $t=-\frac{n}{\chi}$).
\vskip .5cm
\noindent In the first part of this paper we present a stability
 analysis of the travelling 
front (eq. \eqref{eq:twoa}, \eqref{eq:twob})(see~\cite{Zinner1,Zinner2} for an early work demonstrating the existence of stable travelling wave solutions for the Nagumo model), arriving at an
 estimate of a certain `stability
 multiplier' $\rho$ (see below)
 which is to be less than 
unity for the front solution
 to be stable, and show 
numerically that conclusions
 arrived at on the basis of this
 multiplier are indeed conformed
 to in the actual time 
evolution of the FN system
 under consideration. The
 presence of the Heaviside
 function in the model makes
 the linearised evolution
 equation singular, involving
 delta functions at the transition 
points referred to above, and 
so a straightforward calculation
 of the linear growth rates cannot 
be attempted in the model. 
\vskip .5cm
\noindent In section 2, we circumvent this difficulty by
looking more closely at the evolution of perturbations imposed on 
\eqref{eq:twoa}, \eqref{eq:twob}, and estimate the stability multiplier, thereby
arriving at the conclusion that the travelling front solution obtained in~\cite{ref2}
 is stable.
 \vskip .5cm
 \noindent In section 3 we indicate that the system \eqref{eq:onea}, \eqref{eq:oneb}
 involves a certain symmetry, whereby an `anti-kink' solution is associated with a kink solution for any given $w$. We then modify the system \eqref{eq:onea}, \eqref{eq:oneb} by including an equation representing the slow evolution of the recovery variable $w$, and perform a leading order singular perturbation analysis whereby a kink and a corresponding anti-kink solution are pieced together to yield a travelling {\it pulse} solution of the modified system. We also indicate how a {\it pulse train} solution involving an infinite periodic array of uniformly propagating pulses can be obtained in the model.
 \vskip .5cm
 \noindent Section 4 is devoted to concluding remarks, with brief mention of a future communication relating to the stability of the pulse solution.
 \vskip .5cm
 \section{Stability of the travelling front: the stability multiplier}
 \noindent
 As mentioned above, a straightforward linearisation around the travelling front solution presents technical problems involving the appearance of delta functions that make the linearised equations more singular than they actually are. Instead, if we look at the time evolution of a perturbation over the front solution in accordance with (eq. \eqref{eq:onea}, \eqref{eq:oneb}), we find that this involves a succession of 
intervals of linear evolution,
 punctuated by `kicks' at the
 transition points wherein the 
evolution involves a difference
 between two step functions -
 one with and the other without 
the perturbation. For small
 perturbations one can then
 make a convenient 
approximation giving the
 evolution of the 
perturbation through the
 kick.
\vskip .5cm
\noindent More precisely, 
denoting by $\eta_{n}(t)$
  the perturbation over the
 travelling kink soln. $u_{n}(t)$ (eq. \eqref{eq:twoa}, \eqref{eq:twob}),
 the time evolution of 
$\eta_{n}$ is given by 
\begin{eqnarray}
\frac{d(\eta_{n}(t))}{dt}=
D(\eta _{n+1}(t) -
2\eta _{n}(t)+
 \eta_{n-1}(t))- \eta_{n}(t)+
\Theta(u_{n}(t)
+\eta_{n}(t)-a)
-\Theta(u_{n}(t)-a).\label{eq:six}
\end{eqnarray}
For given $n$, $u_{n}(t)$
  crosses the threshold
 $a$ at $t=t_{n}\equiv -\frac{n}{\chi}$.
 Thus, assuming, for instance, $\eta_{n}(t)$
 to be positive, we have
\begin{subequations}
\begin{eqnarray}
\Theta(u_{n}+\eta_{n}-a)-
\Theta(u_{n}-a)=1 \label{eq:sevena}
\end{eqnarray}
\noindent
for
\begin{eqnarray}
 a-\eta_{n} < u_{n}(t) < a, \label{eq:sevenb}
\end{eqnarray}
\end{subequations}
\noindent and zero otherwise. 
\vskip .5cm
\noindent
In the following, the time 
interval during which this
inequality holds will be termed
 a 'kick'. As will be seen below,
 $\eta_{n}$ increases 
monotonically during the kick
 and hence the kick begins when
 
\begin{eqnarray}
u_{n}(t)=a-\eta_{n}^{-}, \label{eq:eight}
\end{eqnarray}
\noindent $\eta_{n}^{-}$ being the 
value of $\eta_{n}$ just before
 the kick. 
\vskip .5cm
\noindent Now, for $t$ sufficiently close
 to but less than $t_{n}$, 
\begin{eqnarray}
u_{n}(t)=a+{\dot{u}}_n(t_{n}^{-})(t-t_{n})+
O(|t-t_{n}|^2), \label{eq:nine}
\end{eqnarray}
where ${\dot{u}}_n(t_{n}^{-})$ is used by recalling that ${\dot{u}}_n$ is actually discontinuous
 at $t=t_{n}$. Thus, for $\eta_n$
 sufficiently small,
\begin{subequations} 
\begin{eqnarray}
u_{n}(t)=a-\eta_{n}^{-} \label{eq:tena}
\end{eqnarray}
\noindent for
\begin{eqnarray}
t \approx t_{n}-
\frac{\eta_{n}^
{-}}{{\dot{u}}_n(t_{n}^{-})}, \label{eq:tenb}  
\end{eqnarray}
\end{subequations}
\noindent where we have assumed that 
${\dot{u}}_n(t_{n}^{-})$ is not so 
small as to make the term 
$O(|t-t_{n}|^{2})$ relevant in \eqref{eq:nine}.
\vskip .5cm
\noindent Thus, the time interval during which 
($\Theta(u_{n}+\eta_{n}-a)-
\Theta(u_{n}-a)$) differs from zero,
 ranges from  ($t_{n}-\frac{\eta_{n}^
{-}}{\dot{u}_{n}(t_{n}^{-})}$) to $t_{n}$. 
This time interval we have 
designated
 above as a `kick'. The time evolution
 of $\eta_{n}(t)$, $(n=0,\pm 1, 
\pm 2,\ldots)$ is then one involving
 a series of kicks, punctuated by
 intervals where the difference
 of $\Theta$-functions in \eqref{eq:six}
 is zero, during which we have 
\begin{eqnarray}
\frac{d\eta_{n}}{dt}-
D((\eta _{n+1}(t) -
2(\eta _{n}(t)+
\eta_{n-1}(t))+ \eta_{n}(t)=0. \label{eq:eleven}
\end{eqnarray}
\vskip .5cm
\noindent For sufficiently small $\eta_{n}$'s
 the kicks are short-lived, and of an 
impulsive nature. Thus, during the 
$n$th
kick we have, to a good degree of 
approximation,
\begin{equation}
\frac{d \eta_{n}}{dt}=1, \label{eq:twelve} 
\end{equation}
\noindent while the other $\eta_{m}$'s 
remain
 almost unaffected during this short
 interval. In between kicks, the
 perturbation evolves according to \eqref{eq:eleven}. 
Equation \eqref{eq:twelve} tells us that $\eta_{n}$ 
indeed 
increases monotonically during 
the kick, and the values of 
$\eta_{n}$ 
just before and just after the 
$n$th kick 
are related as
\begin{eqnarray}
\eta_{n}^{+}=\eta_{n}^{-}+\tau_{n},\label{eq:thirteen}
\end{eqnarray}
\noindent where $\tau_{n}$ is the duration 
of the kick.
\vskip .5cm
\noindent Finally, using \eqref{eq:tenb} one gets,
\begin{eqnarray}
\eta_{n}^{+}=\left(1+\frac{1}
{\dot{u}_{n}(t_{n}^{-})}\right)\eta_{n}^{-}, \label{eq:fourteen}
\end{eqnarray}
\noindent or, using the profile function 
$g(\zeta)$ 
and noting that 
\begin{eqnarray}
\dot{u_n}(t_n^{-})=
\chi g'(\zeta)|_{\zeta=0^{-}}, \label{eq:fifteen}
\end{eqnarray}
\noindent we obtain, 
\begin{equation}
\eta_{n}^{+}=\left(1+\frac{1}
{\chi g'(0^{-})}\right)\eta_{n}^{-}. \label{eq:sixteen}
\end{equation}
We note in passing that a 
straightforward linearisation 
through the replacement of 
($\Theta(u_{n}+\eta_{n}-a)-
\Theta(u_{n}-a)$) by 
$\delta(u_{n}-a)\eta_{n}$ would 
give us
\begin{eqnarray}
\eta_{n}^{+}=e^{\frac{1}
{\chi g'(0^{-})}}\eta_{n}^{-}, \label{eq:seventeen}
\end{eqnarray}
\noindent and would make the problem 
more singular then it actually is. 
In other words, even for small 
$\eta_{n}$,
 one has to take into account the
 non-linearity during the interval
 occupied by the kick, while retaining only the linear term in the {\it duration} of the kick.
\vskip .5cm
\noindent In a similar manner, if $\eta_{n}^{-}$ 
happens to be negative, one finds
\begin{eqnarray}
\eta_{n}^{+}=\frac{1}{1-
\frac{1}{\chi g'(0^{+})}}\eta_{n}^{-}. \label{eq:eighteen}
\end{eqnarray}
\vskip .5cm
\noindent The difference between  
\eqref{eq:sixteen} and \eqref{eq:eighteen} is 
once again a reflection
 of the fact that the non-linearity is relevant 
 during the interval of the kick.
 \vskip .5cm
 \noindent In the following we present
 an approximate stability
 analysis by estimating what
 we have termed the `stability multiplier'
 and our conclusions are
 essentially independent
 of whether we use \eqref{eq:sixteen} or \eqref{eq:eighteen}
 for the amplifying effect of the
 kick on the perturbation. For
 the sake of simplicity we use
 below eq. \eqref{eq:sixteen} in so far as the
 effect of the kick is concerned. 
\vskip .5cm
\noindent We now look into eq. \eqref{eq:eleven} to see
 what happens to the perturbation
 in between the kicks. This is a
 system of linear equations and
 can be easily integrated for a
 time interval $\tau=\frac{1}{\chi}$
 from, say, the end of one kick
 and the begining of the next:
\begin{eqnarray}
\eta_{n}(t+\tau)=
e^{-(2D+1)\tau}
\sum_mI_{n-m}(2D\tau)
\eta_{m}(t), \label{eq:nineteen}
\end{eqnarray}
\noindent where $I_{l}$ stands for the
 Bessel function of order $l$
 with imaginary argument:
\begin{eqnarray}
I_{l}(u)=\frac{1}{2\pi}\int_{0}^
{2\pi}e^{u cos\theta}e^{il\theta}d\theta. \label{eq:twenty}
\end{eqnarray}
\vskip .5cm
\noindent One can also check that 
evolution through \eqref{eq:eleven} leads
 to an over-all damping of the
 perturbation as expected and,
 for instance, obtain the following
 bound,
\begin{eqnarray}
\sum_n |\eta_{n}(t+\tau)|^{2} 
\leq e^{-2\tau}\sum_n |
\eta_{n}(t)|^{2}. \label{eq:twentyone}
\end{eqnarray}
\vskip .5cm
\noindent Indeed, all the eigenvalues of the
 linear problem are real and
 negative, giving us the result \eqref{eq:twentyone}.
 \vskip .5cm
\noindent We now piece together the
 results obtained above. Starting
 from time, say $0^{+}$, just
 after the kick at the site $n=0$
 (recall that a kick affects the
 perturbation at one site only,
 leaving unchanged the other 
sites), the perturbation decays
 through linear evolution till the
 next kick arrives at $t_{-1}=
\frac{1}{\chi}$. There follows the
 impulsive action of the kick,
 amplifying the perturbation at 
site $n=-1$ by the factor 
on the right hand side of \eqref{eq:sixteen}, 
leaving the perturbation at the 
other sites unchanged. The 
process is repeated thereafter,
 the site of action of the kick
 being shifted successively 
by one lattice site. 
\vskip .5cm
\noindent Note that, with
 the shift of the site of action
 of the kick the front itself, 
represented by \eqref{eq:twoa}, \eqref{eq:twob}, moves 
through one lattice site during
 the linear evolution. Since 
the kick at $n=0$ occurs at 
$t_{0}=0$ when the front is 
also located at $n=0$, we 
conclude that {\it each kick amplifies
 the perturbation precisely at
 that lattice site where the
 front is located at 
that instant}. Since the linear
 evolution uniformly damps out the perturbations, we 
see that the most crucial
 factor in respect of the time
 evolution of the perturbation
 resides in the answer to the
 question: what happens to 
the perturbation {\it at the site of
 location of the front} as the 
latter propagates along the 
lattice ? Since the perturbations
 at the other lattice sites are not
 affected by the kicks, they 
are damped out.
\vskip .5cm
\noindent Thus, focussing on the
 perturbation at the location
 of the front (we call it the 
`significant perturbation')
 as it gets shifted from site
 to site, we calculate the
 factor through which it 
gets amplified during the
 interval of one kick and 
the subsequent linear evolution
 till the arrival of the next kick,
 and call it the {\it stability multiplier}.
 The latter is made up of two
 factors, of which one is given
 by eq. \eqref{eq:sixteen}. The other factor is
 to be obtained from eq. \eqref{eq:nineteen} as
 the effect of the linear evolution
 on the significant perturbation.
\vskip .5cm
\noindent One observes, for instance, that
 at the end of the time interval
 (of length $\tau =\frac{1}{\chi}$)
 from $t_{0}^{+}$ (end of kick at 
site $n=0$) up to $t_{-1}^{-}$ 
(begining of kick at site $n=-1$)
 the perturbation at site $n=-1$
 (the significant perturbation at
 time $t_{-1}$) depends linearly 
on the perturbations at the 
various sites at time $t_{0}^{+}$
 (i.e. the time of the previous 
kick). Among the latter, the one
 at site $n=0$ (i.e. the significant
 perturbation at time $t_{0}$) is of
 largest magnitude and happens
 to have the maximum effect on
 the perturbation at site $n=-1$
 at time $t_{-1}=\tau=\frac{1}{\chi}$
 because of its proximity.
 \vskip .5cm
 \noindent Thus, in order to have an order
 of magnitude estimate of the
 stability multiplier, we make the
 simplification of ignoring all
 perturbations on the right 
hand side of eq. \eqref{eq:nineteen} excepting
 the one with $m=-1$ which
 gives, for instance,
\begin{eqnarray}
\eta_{-1}=e^{-(2D+
1)\tau}I_{1}(2D\tau)\eta_{0}(0^{+}). \label{eq:twentytwo}
\end{eqnarray}
\vskip .5cm
\noindent
Noting that $\eta_{0}$ and
$\eta_{-1}$ are the significant 
perturbations at $t=0$ and $t=\tau$ 
respectively, we conclude that the 
second factor in the stability multiplier 
arising due to the linear evolution is 
$e^{\frac{-(2D+
1)}{\chi}}I_{1}(\frac{2D}{\chi})$.
\vskip .5cm
\noindent Thus, finally we arrive at the required 
estimate of the stability multiplier
\begin{subequations}
\begin{eqnarray}
\rho &=& \left(1+
\frac{1}{\chi g'(0^{-})}\right)
e^{\frac{-(2D+
1)}{\chi}}I_{1}(\frac{2D}{\chi}) \label{eq:twentythreea}
\end{eqnarray}
\begin{eqnarray}
= \frac{1}{2\pi}\left(1+
\frac{1}{\chi g'(0^{-})}\right)
e^{\frac{-(2D+
1)}{\chi}}\int_{0}^{2\pi}e^
{\frac{2D}{\chi}cos\theta}
cos\theta d\theta. \label{eq:twentythreeb}
\end{eqnarray}
\end{subequations}
\vskip .5cm
\noindent Making use of the multiplier $\rho$, we 
arrive at the stability criterion,
\begin{eqnarray}
\rho < 1. \label{eq:twentyfour}
\end{eqnarray}
\vskip .5cm
\noindent In reality, eq. \eqref{eq:twentyfour} is not an
 exact criterion because 
(i) eq. \eqref{eq:sixteen} is not an exact 
amplification factor due
 to a kick since it is 
valid for only one class of 
perturbations; 
(ii) it is based on an approach
 that looks at the evolution of
 the significant perturbation alone; and, 
(iii) so far as the effect of the
 linear evolution on the significant
 perturbation is concerned, eq. \eqref{eq:twentytwo} overlooks perturbations at all
 sites excepting the significant
 one at the previous kick.
\vskip .5cm
\noindent Still, one can look upon eq.
 \eqref{eq:twentythreea} as an effective stability
 criterion, that can be used as
 a convenient guide in assessing
 the linear stability of the 
propagating front solution (eq. \eqref{eq:twoa}, \eqref{eq:twob}) 
of the FN system (eq. \eqref{eq:onea}, \eqref{eq:oneb}) under
 consideration. As we indicate
 below, numerical computations
 conform quantitatively to 
conclusions arrived at from eq. \eqref{eq:twentythreea}.
\vskip .5cm
\noindent While accepting eq. \eqref{eq:twentythreea} as a
 stability criterion one, however,
 has to make a couple of 
qualifications relation to its
 validity in the limits $\chi 
\rightarrow 0$
 (pinning limit) and $\chi
 \rightarrow 
\infty$ (zero threshold, see~\cite{ref2}). 
Indeed, making use of
 results in~\cite{ref2} and of the 
asymptotic properties of
 Bessel functions,
one finds, in the pinning 
limit $\chi=0$, 
\begin{subequations}
\begin{eqnarray}
1+\frac{1}{\chi g'(0^{-})} \rightarrow 
\frac{\sqrt{4\pi D}}{\sqrt{\chi}}
e^{\frac{1}{\chi}}, \label{eq:twentyfivea}
\end{eqnarray}
\begin{eqnarray}
e^{-\frac{2D+
1}{\chi}}I_1(\frac{2D}{\chi})
\rightarrow \frac{\sqrt{\chi}}
{\sqrt{4\pi D}}e^{-\frac{1}{\chi}}, \label{eq:twentyfiveb}
\end{eqnarray}
\end{subequations}
\noindent i.e., in the pinning limit, we have,
\begin{eqnarray}
\rho \rightarrow 1. \label{eq:twentysix}
\end{eqnarray}
\noindent On the other hand, in the limit  $\chi \rightarrow \infty$, one similarly has,
\begin{subequations}
\begin{eqnarray}
1+\frac{1}{\chi g'(0^{-})} \rightarrow \frac{\chi}{D}, \label{eq:twentysevena}
\end{eqnarray}
\begin{eqnarray}
e^{-\frac{2D+
1}{\chi}}I_1(\frac{2D}{\chi})
\rightarrow \frac{D}{\chi}, \label{eq:twentysevenb}
\end{eqnarray}
\noindent i.e., once again,
\begin{eqnarray}
\rho \rightarrow 1. \label{eq:twentysevenc}
\end{eqnarray}
\end{subequations}
\vskip .5cm
\noindent Thus, on the face of it, our
 theory makes no definite
 predictions regarding 
stability in  the pinning limit as
 also the limit of of infinitely fast kinks.  
 \vskip .5cm
 \noindent However, in both these 
limits, one notes that the
 amplification factor due 
to the kick given by \eqref{eq:sixteen} 
goes to infinity. This, in fact, is 
an overestimate since for 
$g'(0^{-}) \rightarrow 0$, 
the second term in \eqref{eq:nine} becomes
 vanishingly small and the 
third term, involving 
$g''(0^{-})$ is to be taken
 into account, and thus the 
expression for the kick
 duration
\begin{eqnarray}
\tau_{n}=\frac{\eta_{n}^{-}}
{\chi g'(0^{-})}, \label{eq:twentyeight}
\end{eqnarray}
\noindent is an over-estimation. 
 On the other hand, the damping
 factor due to linear evolution does
 go to zero in both these limits,
 and hence stability is recovered for both $\chi\rightarrow 0$ and $\chi\rightarrow\infty$.
 \vskip .5cm
 \noindent On the other hand, for front 
speeds neither too small nor
 too large, expression \eqref{eq:twentythreea} 
can be taken as a correct 
order-of-magnitude estimate.
 Using the front solution \eqref{eq:twoa}, \eqref{eq:twob}, one
 can compute $\rho$ for given 
values of 
parameters $a$ and $D$ 
characterising the system.   

\begin{figure}[hb]
\centering
\includegraphics[height=6cm,width=9cm]{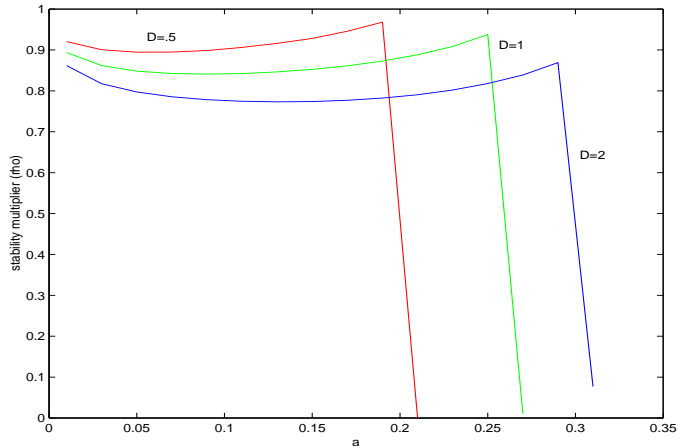}
\caption{\label{cap:f2}Stability multiplier $\rho$ as a function of the threshold parameter $a$ for various different values of $D$. Fall of $\rho$ near the limiting value $\tilde{a}~(\chi\rightarrow 0)$ is an artefact (see text).}
\end{figure}

\vskip .5cm
\noindent In fig. \ref{cap:f2}, we show the variation of the stability multiplier $\rho$ with $a$ for
a set of different values of $D$ where we find that $\rho<1$ for the entire range of parameter values except for $a\rightarrow 0$ ($\chi\rightarrow\infty$) when one has $\rho\rightarrow 1$ in accordance with \eqref{eq:twentysevenc} (one finds in this figure that $\rho\rightarrow 0$ in the pinning limit $a\rightarrow \tilde{a}$ in apparent violation of \eqref{eq:twentysix}; however, this is an artefact because the program used to evaluate \eqref{eq:twentyfivea} yields an underestimation, failing to reproduce the exponential divergence). As already mentioned, \eqref{eq:twentysix}, \eqref{eq:twentysevenc} are overestimations in relation to the actual value of $\rho$ in these two limits. In other words, the travelling front solution obtained in our model is stable for all the parameter values for which it exists. This conclusion is confirmed from fig. \ref{cap:f3} where we show the variation of a related multiplier $\tilde{\rho}$ (see below) with $a$,
\begin{figure}[hb]
\centering
\includegraphics[height=6cm,width=9cm]{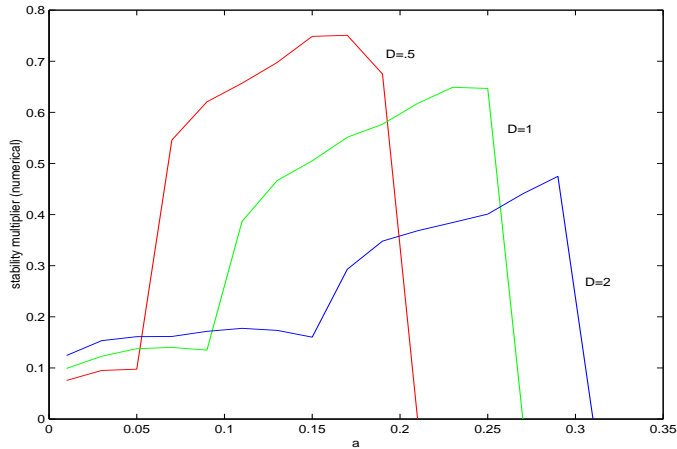}
\caption{\label{cap:f3}Stability multiplier $\tilde{\rho}$ obtained from numerical integration of the Nagumo system (see text) as a function of $a$ for different values of $D$; compare with fig. \ref{cap:f2}. }
\end{figure}
again for a set of different values of $D$. We obtain $\tilde{\rho}$ as follows: we impose a perturbation $\eta_0$ on the travelling front located at the site $n=0$ at time $t=0$, and allow the perturbed system to evolve for a time $\tau=\frac{1}{\chi}$, following the evolution through numerical integration. At the end of this interval, we look at the perturbation $\eta_{-1}$ at site $n=-1$. The ratio $\frac{\eta_{-1}}{\eta_0}$ is then defined as $\tilde{\rho}$, which is thus a numerical estimate for the stability multiplier $\rho$ determined from the actual time evolution of the system under consideration.
\vskip .5cm
\noindent A comparison of figures \ref{cap:f2}, \ref{cap:f3} shows that the variations of $\rho$, $\tilde{\rho}$ are similar in nature (excepting for small $a$, see above), indicating that our theoretical estimate gives qualitatively correct predictions relating to the stability of the propagating front, and the front solution is indeed stable in its entire range of existence.
\begin{figure}[hb]
\centering
\includegraphics[height=6cm,width=9cm]{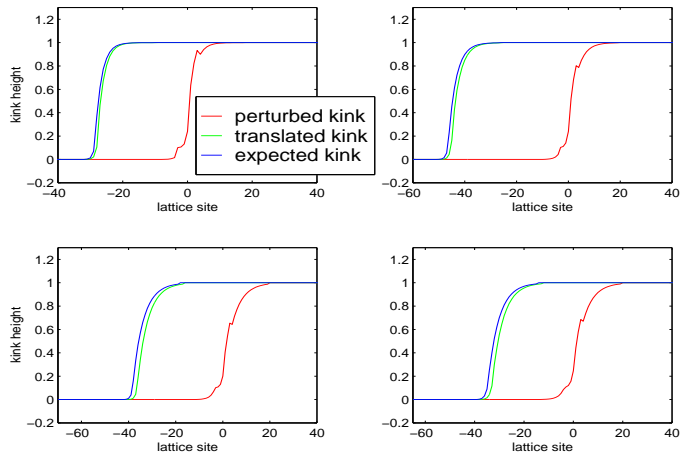}
\caption{\label{cap:f4}Numerical integration of the Nagumo system for an initial kink solution together with a perturbation; clockwise from top left (a) $D=1, a=0.1382, \tau=20$, (b) $D=2, a=0.1, \tau=20$, (c) $D=3, a=0.1445, \tau=10$, (d) $D=4, a=0.2272, \tau=10$; in each frame, the profile obtained through numerical integration for a time $\tau$ (`translated kink') has been compared with the profile obtained from the theoretically obtained solution with $t=\tau$ (`expected kink'). }
\end{figure}
\vskip .5cm
\noindent Fig. \ref{cap:f4} presents results of numerical integration of \eqref{eq:onea},
\eqref{eq:oneb} for more general perturbations. We impose over the kink profile a perturbation spread over a few sites around $n=0$ at $t=0$ (recall that this includes what we have termed the significant perturbation) and look at the front profile after an appropriate time interval to see what has happened to the perturbation. It is found that, for all values of the parameters $a$, $D$ for which the integration is performed (only a few representative ones among these are shown in fig. \ref{cap:f4}) the perturbation dies down with time.
\vskip .5cm
\noindent In summary, we conclude that the stability multiplier $\rho$ obtained above gives a reliable indication of the temporal stability of the travelling front solution and that the latter is, in all likelihood, stable for all values of the parameters $a$, $D$ for which it exists.
\section{The propagating anti-kink, pulse, and pulse-train solutions}
\subsection{The anti-kink}
\begin{figure}[hb]
\centering
\includegraphics[height=6cm,width=9cm]{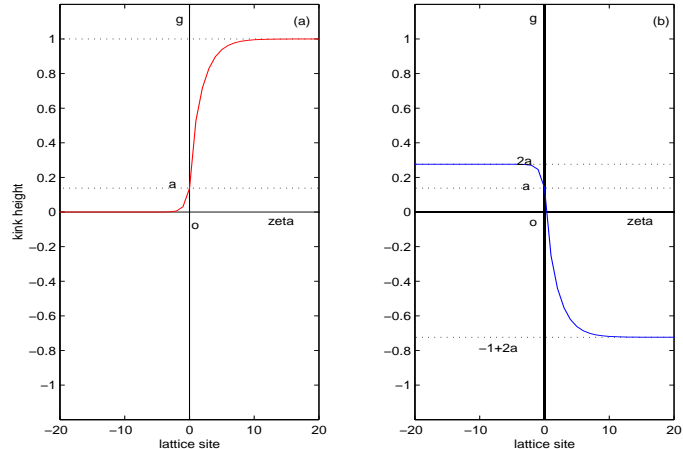}
\caption{\label{cap:f5} The (a) kink and (b) anti-kink profiles as function of $\zeta$ (corresponding to integers representing lattice sites at $t=0$) showing level changes for given $D(=1)$ and $a(=0.1382)$; introduction of the slow dynamics connects up the kink and the anti-kink into a travelling pulse (see text). }
\end{figure}

\noindent We begin by noting that equations \eqref{eq:onea}, \eqref{eq:oneb} possess a symmetry $w_n\rightarrow 1-2a-w_n,~u_n\rightarrow 2a-u_n$, as a result of which there is associated, with a travelling front or kink solution, a travelling `anti-kink' propagating with the same speed $\chi$. The latter is obtained from the former, equations \eqref{eq:twoa}, \eqref{eq:twob}, by making the above transformation. Fig. \ref{cap:f5} depicts schematically the level change in $u_n$ (for a given lattice site $n$) for the kink and the
corresponding anti-kink solution, the level change for the latter being shown in such a manner that both the kink and the anti-kink propagate in the same direction along the lattice.
\vskip .5cm
\subsection{Fast and slow dynamics: the travelling pulse solution}
\begin{figure}[hb]
\centering
\includegraphics[height=6cm,width=9cm]{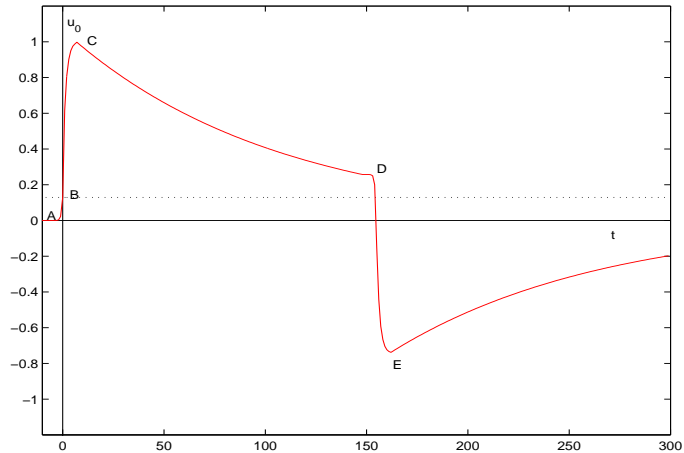}
\caption{\label{cap:f6}$u_0$ as a function of $t$ for a typical pulse solution showing schematically the points of matching between the fast and slow dynamics; the choice $n=0$ for the lattice site is arbitrary. }
\end{figure}

\begin{figure}[hb]
\centering
\includegraphics[height=6cm,width=9cm]{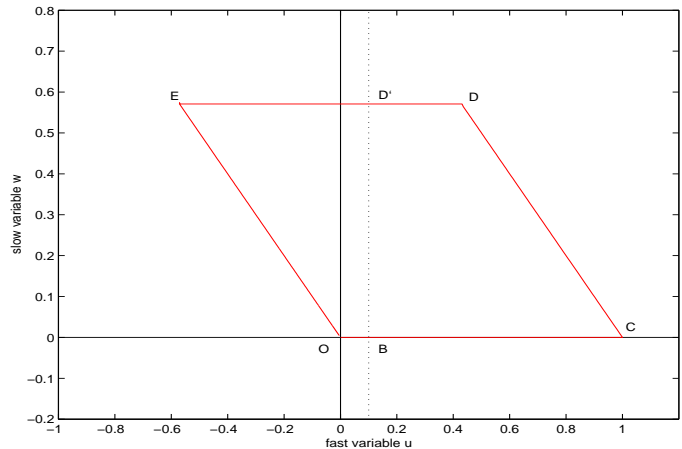}
\caption{\label{cap:f7}$u_0$-$w_0$ (the choice $n=0$ for the lattice site is arbitrary) phase diagram for the pulse solution showing points of matching as also the points where $u_0$ crosses the threshold $a$; the origin (O) is the asymptotic point for the orbit for $t\rightarrow\pm\infty$. }
\end{figure}

\noindent We now modify our system \eqref{eq:onea}, \eqref{eq:oneb} in such a way that there occurs a slow change in the recovery variable $w_n$ as a result of which, after the level change in $u_n$ in the kink solution (for any given $n$, with $w_n=0$) from a low to a high level (with reference to the threshold $a$), there occurs a slow rise in $w_n$ and a corrsponding slow fall in $u_n$ till there appears a rapid level change in $u_n$ in accordance with the anti-kink solution matching the earlier kink solution, now with $w_n=1-2a$. Thereafter, there occurs a slow change in $w_n$ and $u_n$ whereby both return to the resting value $0$. Such alteration of fast and slow dynamics is actually observed in excitable media of interest, and provides the basis for FitzHugh-Nagumo dynamics (see, e.g.,~\cite{Hag}) as a prototype model for such systems.
\vskip .5cm
\noindent In an early paper, Conley~\cite{ref3} gave a geometric argument based on a singular perturbation approach where he established the existence of a homoclinic orbit representing a pulse solution (in the context of a 1D continuous excitable medium) on reducing the partial differential equations  to a system of ODE's in terms of an appropriate propagation variable. Such pulse solutions in spatially continuous systems have since been extensively discussed in the literature (see, e.g.,~\cite{ref4,ref5,ref6,Rajagopal}). In particular the question of stability of these pulses, obtained by piecing together a leading front and a trailing rear (respectively the `kink' and the `anti-kink'), has been addressed  in the context of the so-called restitution hypothesis~\cite{ref7,ref8}.
\vskip .5cm
\noindent In the following, we present a leading order singular-perturbation construction, along similar lines, of the pulse solution in a 1D {\it discrete} lattice in the above-mentioned modification of the system \eqref{eq:onea}, \eqref{eq:oneb} obtained by introducing the slow dynamics involving the recovery variable $w_n$. More precisely, the system we now consider is 
\begin{subequations}
\begin{eqnarray}
\frac{d u_{n}}{dt}=D(u_{n+1}
-2{u_n}+u_{n-1})+f(u_n),\label{eq:twentyninea}
\end{eqnarray}
\begin{eqnarray}
\frac{d w_{n}}{dt}=\epsilon u_n, \label{eq:twentynineb}
\end{eqnarray}
\end{subequations}
\noindent where $\epsilon$ is a small recovery parameter setting the time scale of the slow dynamics in terms of which we perform below the leading order singular perturbation calculation.
\vskip .5cm
\noindent We now have two pieces of inner solution describing a rapid level change in $u_n$ (for $w_n=0$ and $w_n=1-2a$ for the kink and the anti-kink respectively) and two other pieces of outer solution involving the slow change of $u_n$ as also of $w_n$. Fig. \ref{cap:f6} shows schematically the variation of $u_0$ with $t$ in which ABC depicts the leading edge and DE the trailing edge while CD and EF correspond to the slow dynamics. The points C, D, and E indicate the points of matching between the inner and outer solutions, which we work out below. Fig. \ref{cap:f7} depicts the pulse solution in the $u_0$-$w_0$ (the choice $n=0$ is arbitrary ) phase plane, where the matching points are indicated once again, together with the points B, D$'$ where $u_0$ crosses the threshold with $w_0\approx 0$ and $w_0\approx 1-2a$ respectively. Note that the pulse rises from $u_0=0$ at $t\rightarrow -\infty$ and finally recovers to $u_0=0$ at $t\rightarrow +\infty$. Thus, from here on, we denote by $g(\zeta)$ the kink solution described in section 1, {\it with the recovery parameter} $w$ {\it set at} $0$ (the same notation will apply for the pulse-train solution as well).

\vskip .5cm
\noindent Following~\cite{ref2} we introduce the propagation variable $\zeta=\chi t+n$, and represent the pulse solution as 
\begin{subequations}
\begin{eqnarray}
u_n(t)=u(\zeta),
\end{eqnarray}
\begin{eqnarray}
w_n(t)=w(\zeta),
\end{eqnarray}
where
\begin{eqnarray}
lim_{\zeta\rightarrow \pm \infty}u(\zeta)=lim_{\zeta\rightarrow \pm \infty}w(\zeta)=0.
\end{eqnarray}
\end{subequations}
\vskip .5cm
\noindent Let the values of the propagation variable at the matching points be denoted by $\zeta_C$, $\zeta_D$, and $\zeta_E$ respectively, which we calculate below in the leading order in $\epsilon$.
\vskip .5cm
\noindent Starting from the kink solution \eqref{eq:twoa}, \eqref{eq:twob}, for $\epsilon=0$, $w=0$, one obtains the leading correction as we switch on a small $\epsilon$:
\begin{eqnarray}
w(\zeta)=\frac{\epsilon}{\chi}\int_{-\infty}^{\zeta}g(\zeta)d\zeta.
\end{eqnarray}
\noindent Noting that $g(\zeta)\sim 0$ (here and in the following we use the symbol `$\sim$' to denote a leading order approximation in $\epsilon$) for $\zeta$ away from and less than $0$, while $g(\zeta)\sim 1$ for $\zeta$ greater than and away from $0$, we get the asymptotic expression for $w(\zeta)$ for large $\zeta$:
\begin{eqnarray}
w(\zeta)\sim \frac{\epsilon\zeta}{\chi}.
\end{eqnarray}
\noindent With this as the inner solution for $w$ for $0<\zeta<\zeta_C$, and the corresponding inner solution for $u$ as 
\begin{eqnarray}
u(\zeta)\sim g(\zeta),
\end{eqnarray}
\noindent the {\it outer} solution satisfies, from eq. \eqref{eq:twentyninea},
\begin{subequations}
\begin{eqnarray}
u(\zeta)\sim 1-w(\zeta).
\end{eqnarray}
\noindent where, from \eqref{eq:twentynineb}
\begin{eqnarray}
w(\zeta)\sim 1-e^{-\frac{\epsilon}{\chi}\zeta}.\label{eq:slowdyn}
\end{eqnarray}
\end{subequations}
\noindent Matching the inner and outer solutions, we get the value of $\zeta_C$ from
\begin{eqnarray}
g(\zeta_C)\sim 1-\frac{\epsilon\zeta_C}{\chi}.\label{eq:matchC}
\end{eqnarray}
\noindent  It is now necessary to obtain the asymptotic expression for $g(\zeta)$ for large $|\zeta|$ in order to solve for $\zeta_C$ from above. This can be done by noting that, in the asymptotic region, the linear approximation holds for the evolution of $u_n(t)$, i.e., 
\begin{eqnarray}
\chi\frac{dg(\zeta)}{d\zeta}=D(g(\zeta+1)-2g(\zeta)+g(\zeta-1))-g(\zeta)+g_0,
\end{eqnarray}
\noindent where $g_0=1$ (resp. $0$) for $\zeta\rightarrow +\infty$ (resp. $\zeta\rightarrow -\infty$). Then, assuming 
\begin{eqnarray}
g(\zeta)-g_0\sim \sigma^{|\zeta|}~(say),
\end{eqnarray} 
we have, 
\begin{eqnarray}
\chi ln\sigma = D(\sigma+\sigma^{-1})-(2D+1),\label{eq:sigma}
\end{eqnarray}
\noindent from which one can determine the exponent $\sigma$. As an example, we have the results
\begin{eqnarray}
\sigma=\gamma, ~~for~ \chi\rightarrow 0,
\end{eqnarray}
and
\begin{eqnarray}
\sigma=e^{-\frac{1}{\chi}}, ~~for ~ \chi\rightarrow\infty,
\end{eqnarray}
i.e., 
\begin{eqnarray}
g(\zeta)=g_0+\alpha\gamma^{|\zeta|},~~for~\chi\rightarrow 0,
\end{eqnarray}
\noindent and
\begin{eqnarray}
g(\zeta)=g_0+\beta(e^{{-\frac{1}{\chi}}})^{|\zeta|},~~for ~\chi\rightarrow\infty,
\end{eqnarray}
\noindent where $\alpha$, $\beta$ are constants whose values need not concern us here.
\vskip .5cm
\noindent Thus, in the following, we write $g(\zeta)=g_0+A\sigma^{|\zeta|}$ for $|\zeta|\rightarrow\infty$ for arbitrarily specified $\chi$, $D$, where the constant $A$ is not relevant for our analysis, and where $\sigma$ is to be determined from the transcendental equation \eqref{eq:sigma}.
\vskip .5cm
\noindent Thus, finally, the matching at the point C gives (ref. eq. \eqref{eq:matchC}) 
\begin{eqnarray}
1+A\sigma^{\zeta_C}=1-\frac{\epsilon\zeta_C}{\chi},
\end{eqnarray}
giving, 
\begin{eqnarray}
\zeta_C\sim\frac{ln\epsilon}{ln\sigma}.\label{eq:zetaC}
\end{eqnarray}
\noindent Next, it is easy, from \eqref{eq:slowdyn}, to calculate to leading order in $\epsilon$ the time or $\zeta$-interval from C to D, the latter being the point where $w(\zeta)\sim 1-2a$. One thereby gets
\begin{eqnarray}
\zeta_D-\zeta_C\sim\frac{\chi}{\epsilon}ln(\frac{1}{2a}).\label{eq:zetaD}
\end{eqnarray}
\noindent As expected, the interval ($\zeta_C$) after which the fast dynamics is succeeded by the slow dynamics, is small compared to the interval ($\zeta_D-\zeta_C$) during which the slow dynamics operates, in turn giving way to the fast level change from D to E. The interval corresponding to the latter is obtained as a sum of two sub-intervals : (a) the interval necessary for $u(\zeta)$ to drop from $2a$ to $a$ (point D$'$ in fig. \ref{cap:f7}), and (b) the interval for $u(\zeta)$ to drop from $a$ to $u_\zeta|_E\sim -1+2a$. However, in order to calculate these sub-intervals to the leading order in $\epsilon$, one has to again perform a matching between the outer and the inner solutions, as was done for the point C. One thereby obtains
\begin{eqnarray}
\zeta_E-\zeta_D\sim 2\frac{ln\epsilon}{\ln\sigma}.\label{eq:zetaE}
\end{eqnarray}
\noindent Finally, the slow dynamics takes over from E onwards, and the pulse returns to the level $u=0,~w=0$ over an infinite $\zeta$-interval.
\vskip .5cm
\noindent Summarising, we present below the travelling pulse solution obtained in the leading order of sungular perturbation calculation:
\begin{subequations}
\begin{eqnarray}
-\infty<\zeta<\zeta_C : ~~~u(\zeta)\sim g(\zeta),\label{eq:pulse1}
\end{eqnarray}
\begin{eqnarray}
~~~~~~~~~~~~~~~~~~~w(\zeta)\sim \frac{\epsilon}{\chi}\int_{-\infty}^\zeta g(\zeta)d\zeta,\label{eq:pulse2}
\end{eqnarray}
\begin{eqnarray}
\zeta_C<\zeta<\zeta_D:~~~u(\zeta)\sim e^{-\frac{\epsilon}{\chi}(\zeta-\zeta_C)},\label{eq:pulse3}
\end{eqnarray}
\begin{eqnarray}
~~~~~~~~~~~~~~~~~~~~w(\zeta)\sim 1-e^{-\frac{\epsilon}{\chi}(\zeta-\zeta_C)},\label{eq:pulse4}
\end{eqnarray}
\begin{eqnarray}
\zeta_D<\zeta<\zeta_E:~~~u(\zeta)\sim 2a-g(\zeta-\frac{\zeta_E+\zeta_D}{2}),\label{eq:pulse5}
\end{eqnarray}
\begin{eqnarray}
~~~~~~~~~~~~~~~~~~~~~~w(\zeta)\sim 1-2a,\label{eq:pulse6}
\end{eqnarray}
\begin{eqnarray}
\zeta_E<\zeta<\infty:~~~u(\zeta)\sim -(1-2a)e^{-\frac{\epsilon}{\chi}(\zeta-\zeta_E)},\label{eq:pulse7}
\end{eqnarray}
\begin{eqnarray}
~~~~~~~~~~~~~~~~~~~~~~~w(\zeta)\sim (1-2a)e^{-\frac{\epsilon}{\chi}(\zeta-\zeta_E)},\label{eq:pulse8}
\end{eqnarray}
\end{subequations}
\noindent where $\zeta_C,~\zeta_D,~\zeta_E$ have been obtained above (see equations \eqref{eq:zetaC}, \eqref{eq:zetaD}, \eqref{eq:zetaE}). Note that in this pulse solution the matching of $u(\zeta)$ as also of $w(\zeta)$ at $\zeta_C,~\zeta_D$, and             $\zeta_E$ is accurate only in the leading order in $\epsilon$, and so are the values of $\zeta_C,~\zeta_D,~ \zeta_E$ themselves.
\begin{figure}[hb]
\centering
\includegraphics[height=6cm,width=9cm]{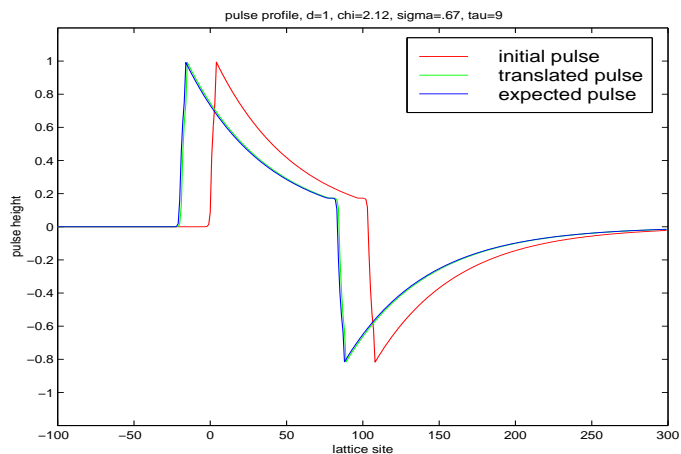}
\caption{\label{cap:f8}An initial profile obtained from the theoretically calculated pulse solution with $t=0$, together with the profile (`translated pulse') obtained from the initial profile on numerical integration of the FitzHugh-Nagumo system for a time lapse $\tau$, and the theoretically calculated profile drawn with $t=\tau$ (`expected pulse'); $D=1, \epsilon=.04, \sigma=0.67, a=0.0864, \tau=9$; the translated and expected pulses coincide in the scale of the figure. }
\end{figure}
\begin{figure}[hb]
\centering
\includegraphics[height=6cm,width=9cm]{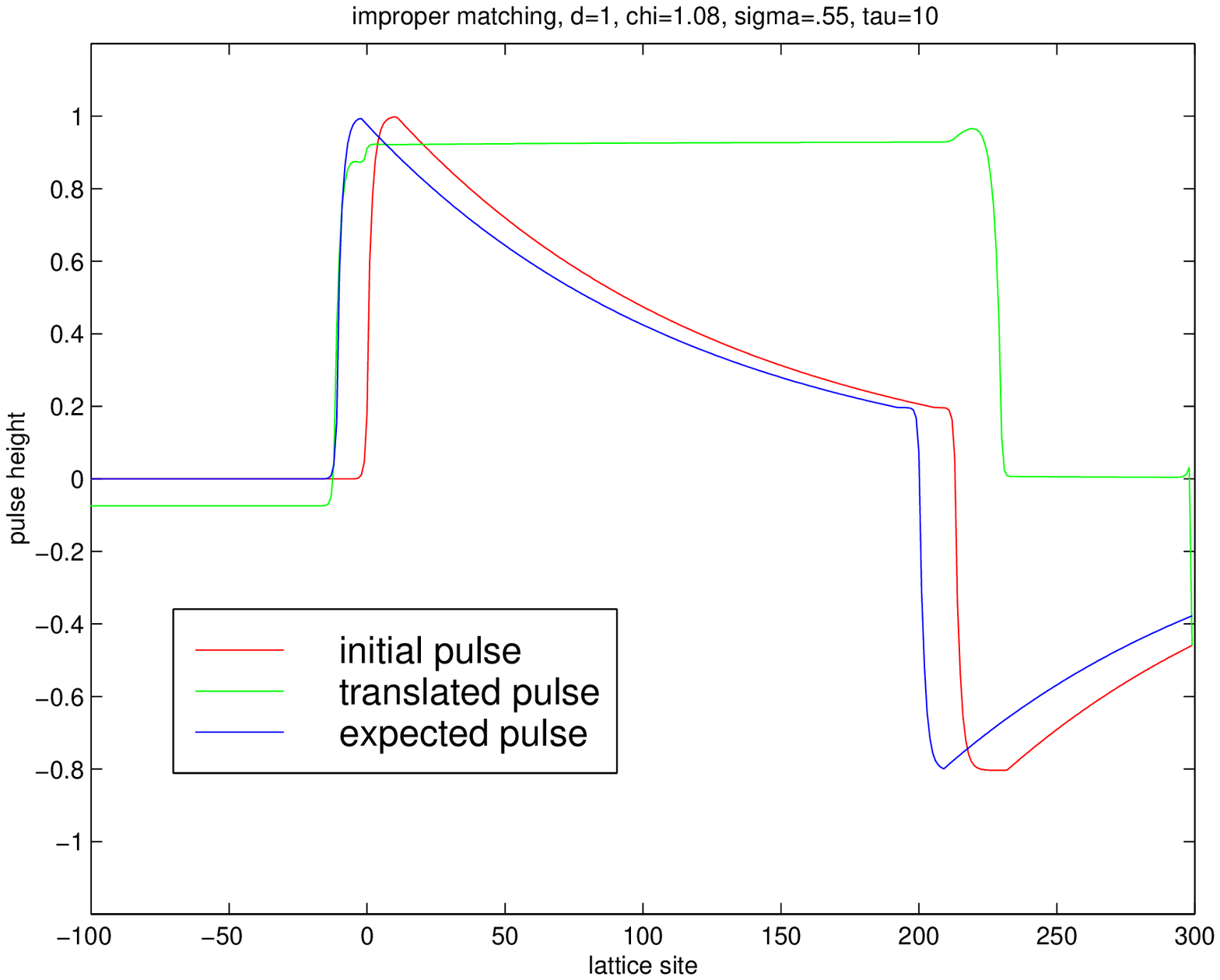}
\caption{\label{cap:f9}Same as fig. \ref{cap:f8}, but with an improper matching between the fast and slow dynamics (see text); $D=1, \epsilon=0.009, \sigma=0.5533, a=0.0982, \tau=10$ . }
\end{figure}
\vskip .5cm
\noindent Fig. \ref{cap:f8} shows an initial ($t=0$) pulse profile constructed in accordance with eq. \eqref{eq:pulse1} -  \eqref{eq:pulse8} as also the profile obtained by numerical integration of \eqref{eq:twentyninea}, \eqref{eq:twentynineb} with this initial condition for a time lapse $\tau$ (see caption). At the same time, it depicts the profile obtained from \eqref{eq:pulse1} -  \eqref{eq:pulse8} by putting $t=\tau$. The agreement of the latter two indicates that \eqref{eq:pulse1} -  \eqref{eq:pulse8} does indeed constitute a travelling pulse solution of \eqref{eq:twentyninea}, \eqref{eq:twentynineb}.
\vskip .5cm
\noindent On the other hand, when a similar exercise is performed (fig. \ref{cap:f9}) with a pulse constructed as in \eqref{eq:pulse1} -  \eqref{eq:pulse8} but with matchings done at $\zeta$-values deviating from those in \eqref{eq:zetaC}- \eqref{eq:zetaE} (we use changed values $\zeta_C'=2\zeta_C,~ \zeta_D'-\zeta_C'=\zeta_D-\zeta_C,~\zeta_E'-\zeta_D'=2(\zeta_E-\zeta_D)$), one observes that the profile at $t=\tau$ differs markedly from that obtained by numerical integration of the initial profile for a time lapse $\tau$, thereby confirming the validity of our singular perturbation construction.
\begin{figure}[hb]
\centering
\includegraphics[height=6cm,width=9cm]{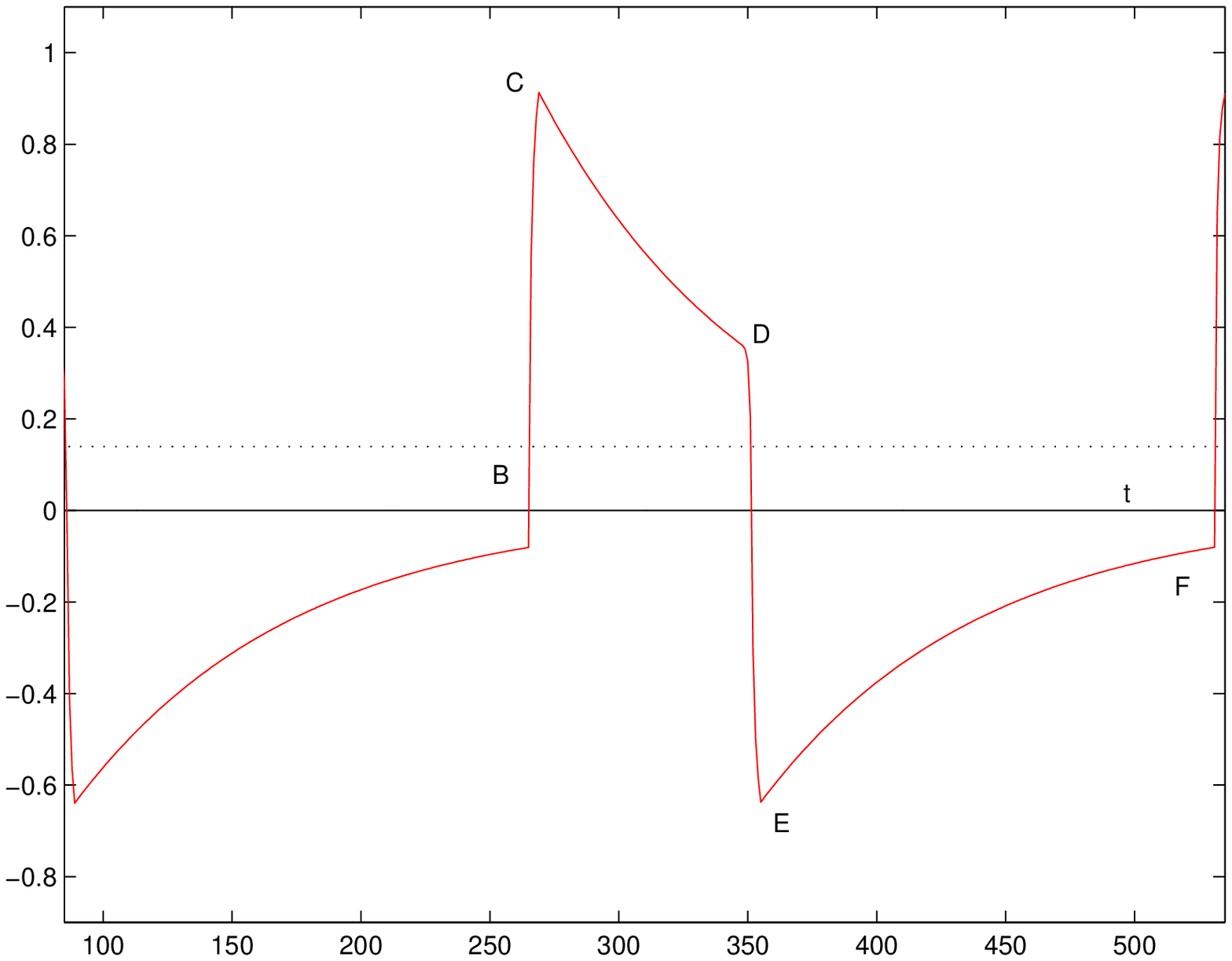}
\caption{\label{cap:f10}$u_0$ as a function of $t$ for a typical pulse-train solution showing schematically the matching points (similar to fig. \ref{cap:f6}); the trailing edge of one pulse joins up with the leading edge of the next pulse, and thus there are an infinite succession of matching points. }
\end{figure}
\begin{figure}[hb]
\centering
\includegraphics[height=6cm,width=9cm]{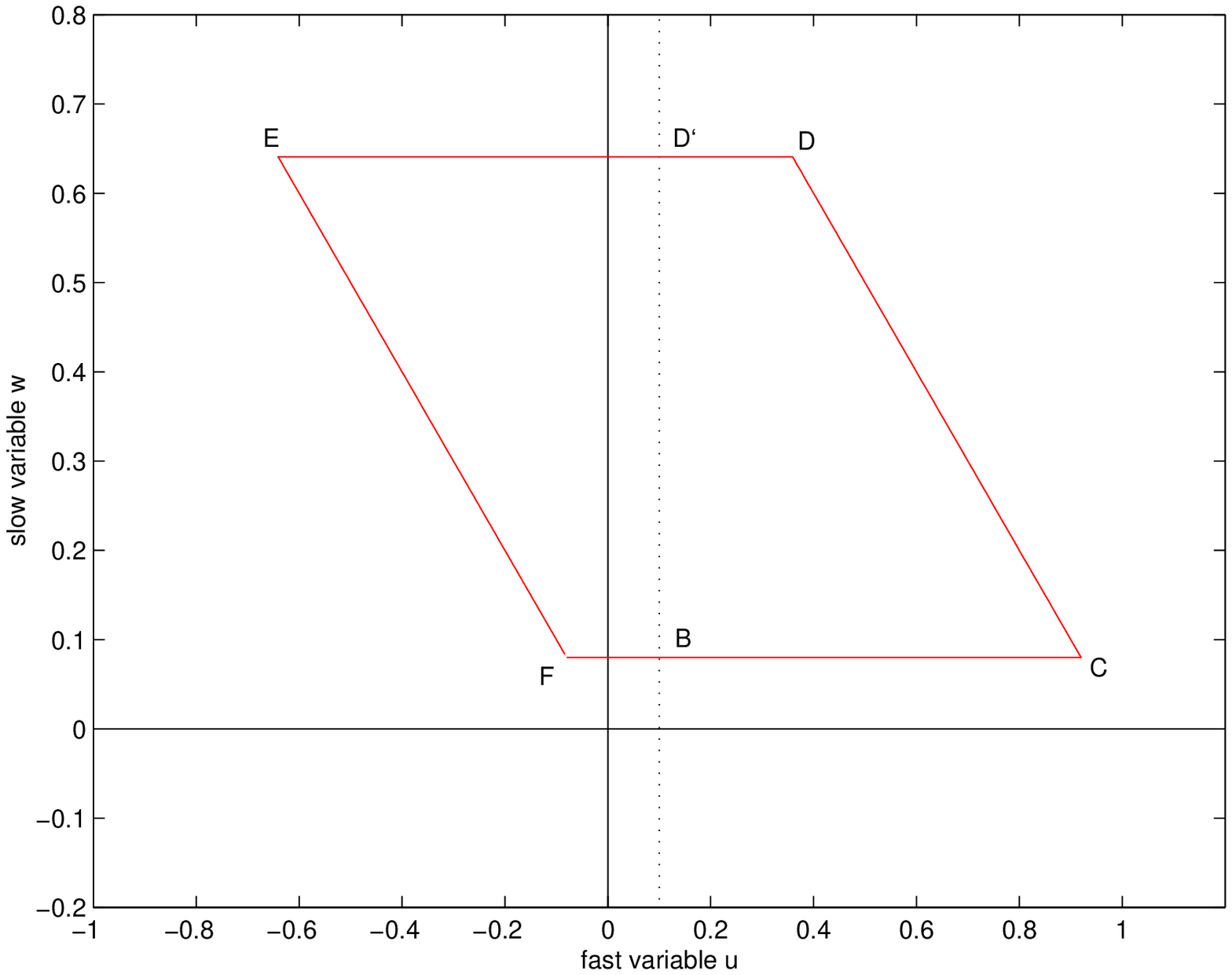}
\caption{\label{cap:f11}$u_0$-$w_0$ phase diagram for a pulse-train solution (similar to fig. \ref{cap:f7}); the closed loop is traversed periodically. }
\end{figure}
\subsection{The periodic pulse-train}
\noindent Simiar principles can be used to construct a one-parameter family of travelling {\it pulse-trains} where a pulse-train involves a periodic succession of pulses travelling along the lattice with a constant speed. Whereas a lone pulse involves a leading edge with a level change (from $u=0$ to $u=1$) at $w=0$ and a trailing edge with a level change (from $u=2a$ to $u=-1+2a$) at $w=1-2a$, the leading edge of each pulse in a pulse train involves a level change at some non-zero value (say, $W$) of the recovery variable $w$. where $W$ is to lie in the range $0<W<\tilde{a}-a$ where the upper limit is obtained by noting that a kink solution with its level change occurring at $w=\tilde{a}-a$ gets pinned in the lattice.
\vskip .5cm
\noindent Fig. \ref{cap:f10} depicts schematically the variation of $u_0$ with $t$ where successive level changes occur with alternating fast and slow dynamics, the points of matching between the inner and outer solutions being shown in a manner similar to fig. 6. 
\begin{figure}[Ht]
\centering
\includegraphics[height=6cm,width=9cm]{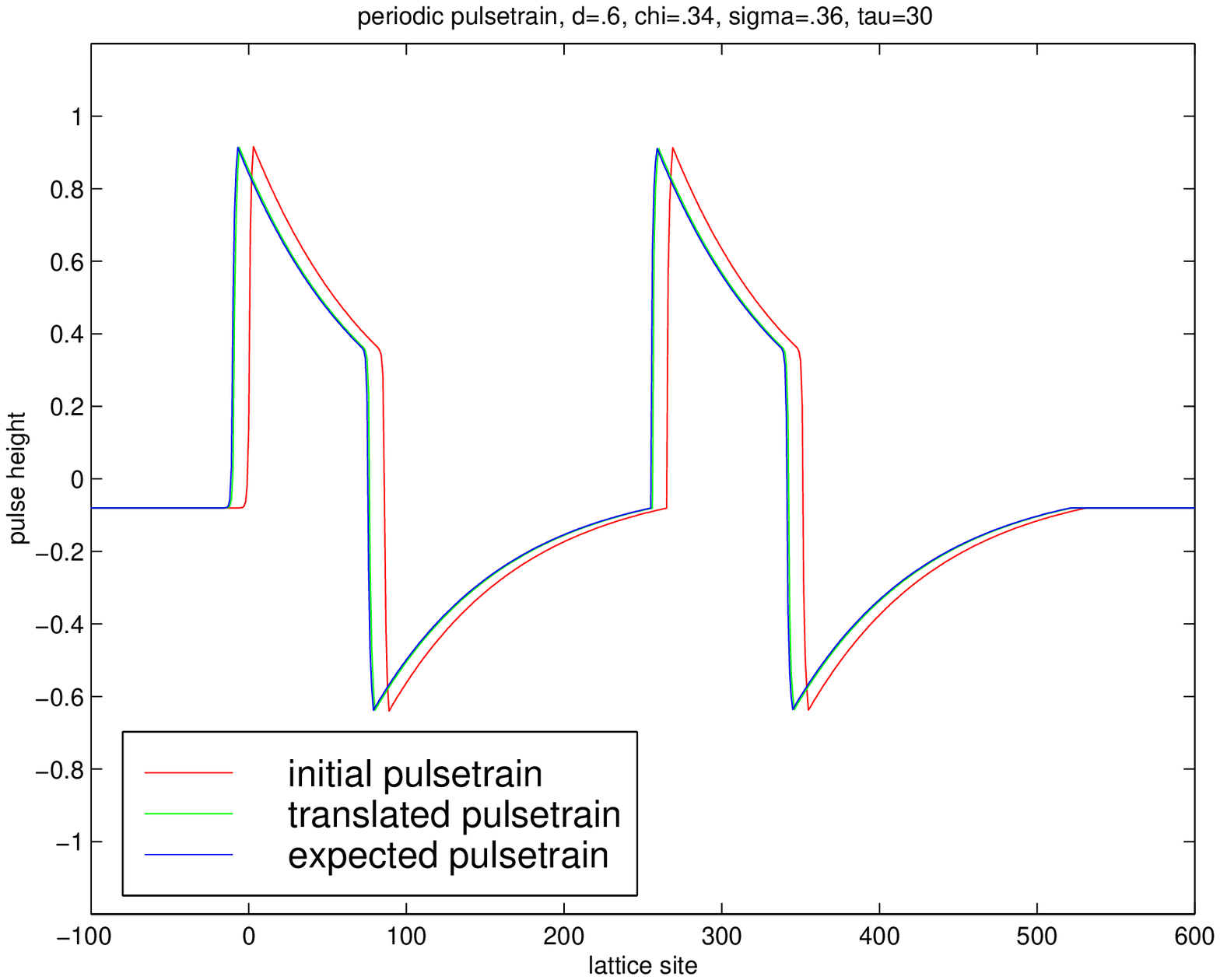}
\caption{\label{cap:f12}Comparison of theoretically calculated and numerically computed pulse-train solution (similar to comparison in fig. \ref{cap:f8} for a lone pulse); $D=0.6, a=0.1395, W=0.08, \epsilon=0.004, \tau=30, \sigma=0.3661$; once again, the `translated' and `expected' pulse trains coincide. }
\end{figure}
Note that the leading edge of each pulse, instead of rising from $u_0=0$, joins up with the trailing edge of the previous pulse in the pulse train and thus, for each pulse in the train there are four matching points (C, D, E, F in fig. \ref{cap:f10}) instead of three (cf. fig. \ref{cap:f6}) for a lone pulse. Fig. \ref{cap:f11} shows the pulse train dynamics in the $u_0-w_0$ phase plane where the closed curve is traversed repeatedly, and the matching
\begin{figure}[Ht]
\centering
\includegraphics[height=6cm,width=9cm]{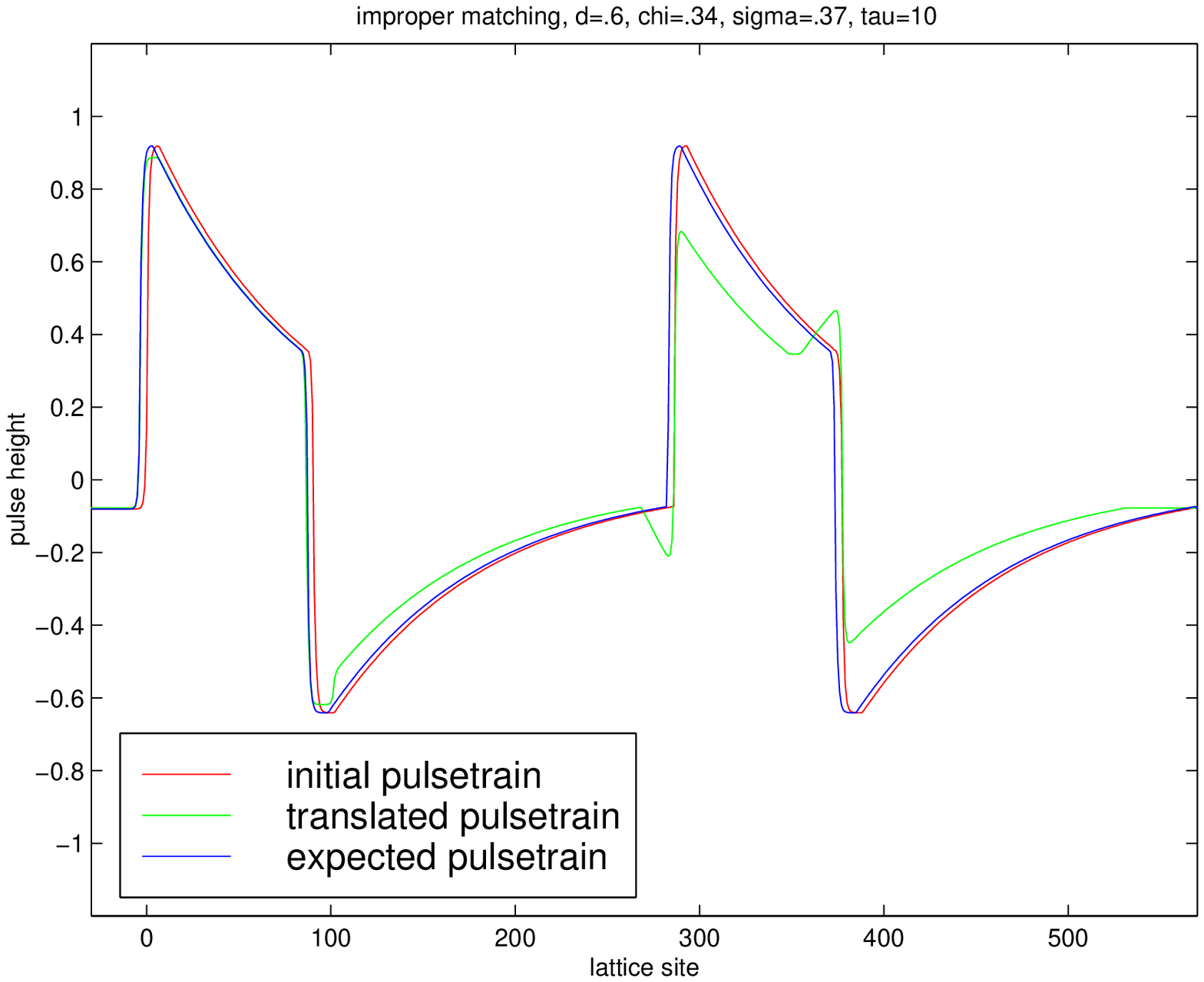}
\caption{\label{cap:f13}Comparison of theoretically calculated and numerically computed pulse-train solution (similar to comparison in fig. \ref{cap:f9} for a lone pulse), with an improper matching between the fast and slow dynamics (see text); parameters same as in fig. \ref{cap:f12}, but with $\tau=10$. }
\end{figure}
 points together with the points of threshold crossing are indicated.
\vskip .5cm
\noindent Applying the matching principles explained for the lone pulse, we obtain the pulse train solution as below (as explained above, we have a 1-parameter family of such solutions characterised by $W$):
\begin{subequations}
\begin{eqnarray}
-\zeta_1<\zeta<\zeta_1 : ~~~u(\zeta)\sim g(\zeta)-W,\label{eq:train1}
\end{eqnarray}
\begin{eqnarray}
~~~~~~~~~~~~~~~~~~~w(\zeta)\sim W, \label{eq:train2}
\end{eqnarray}
\begin{eqnarray}
\zeta_1<\zeta<\zeta_1+\zeta_2:~~~u(\zeta)\sim (1-W) e^{-\frac{\epsilon}{\chi}(\zeta-\zeta_1)},\label{eq:train3}
\end{eqnarray}
\begin{eqnarray}
~~~~~~~~~~~~~~~~~~~~w(\zeta)\sim 1-u(\zeta),\label{eq:train4}
\end{eqnarray}
\begin{eqnarray}
\zeta_1+\zeta_2<\zeta<3\zeta_1+\zeta_2:~~~u(\zeta)\sim 2a-g(\zeta-\zeta_1-2\zeta_2)+W,\label{eq:train5}
\end{eqnarray}
\begin{eqnarray}
~~~~~~~~~~~~~~~~~~~~~~w(\zeta)\sim 1-2a-W,\label{eq:train6}
\end{eqnarray}
\begin{eqnarray}
3\zeta_1+\zeta_2<\zeta<3\zeta_1+\zeta_2+\zeta_3:~~~u(\zeta)\sim -(1-2a-W)e^{-\frac{\epsilon}{\chi}(\zeta-3\zeta_1-\zeta_2)},\label{eq:train7}
\end{eqnarray}
\begin{eqnarray}
~~~~~~~~~~~~~~~~~~~~~~~w(\zeta)\sim -u(\zeta), \label{eq:train8}
\end{eqnarray}
\begin{eqnarray}
u(\zeta+Z)=u(\zeta),\label{eq:train9}
\end{eqnarray}
\begin{eqnarray}
w(\zeta+Z)=w(\zeta),\label{eq:train}
\end{eqnarray}
\end{subequations}
\noindent where
\begin{subequations}
\begin{eqnarray}
\zeta_1\sim\frac{ln\epsilon}{ln\sigma},
\end{eqnarray}
\begin{eqnarray}
\zeta_2\sim \frac{\chi}{\epsilon}ln(\frac{1-W}{2a+W}),
\end{eqnarray}
\begin{eqnarray}
\zeta_3\sim \frac{\chi}{\epsilon}ln(\frac{1-W-2a}{W}),
\end{eqnarray}
and the periodicity is
\begin{eqnarray}
Z=4\zeta_1+\zeta_2+\zeta_3.
\end{eqnarray}
\end{subequations}
\vskip .5cm
\noindent The speed $\chi$ of the pulse-train depends on the initial level $W$ of the recovery variable and is given, in the leading order of $\epsilon$, by the equation
\begin{eqnarray}
\frac{1}{2\pi}\int_0^{2\pi}\frac{d\theta}{(1-\nu cos\theta)(1-e^{-i\theta}e^{-\mu(1-\nu cos\theta)})}=-(2D+1)(a+W-a_0)
\end{eqnarray}
\noindent (cf. eq. (26) of ~\cite{ref2} giving the speed of the kink solution for $W=0$).
\vskip .5cm
\noindent Fig. \ref{cap:f12} shows a pulse-train profile (a) at $t=0$ and (b) at $t=\tau$ (see caption) obtained from \eqref{eq:train1} - \eqref{eq:train}, together with (c) the profile obtained by numerical integration of \eqref{eq:twentyninea}, \eqref{eq:twentynineb} for a time lapse $\tau$. The agreement between (b) and (c) indicates that the pulse-train solution obtained above is indeed a valid one. Additionally, fig. \ref{cap:f13} shows the result of a similar exercise but now with an improper matching between the inner and outer solutions using $\zeta$-intervals as in fig. \ref{cap:f9}. One observes here that the pulse-train profile at $t=\tau$ differs substantially 
from the result of numerical integration from the initial profile.
\section{Concluding remarks}
\noindent In summary, we have, in this paper, established the temporal stability of the travelling front solution \eqref{eq:twoa} - \eqref{eq:nu} of the  discrete reaction-diffusion system \eqref{eq:onea}, \eqref{eq:oneb} by way of estimating the stability multiplier and have constructed the travelling pulse as also a one-parameter family of pulse-train solutions for a system including a slow variation of the recovery parameter, through a leading order singular perturbation analysis. The travelling pulse consists of a leading edge  and a trailing edge corresponding to levels $w=0$ and $w=1-2a$ of the recovery variable, and also intervals of slow dynamics as explained above. A pulse-train, on the other hand, is made up of a periodic succession of pulses with the leading edge of one pulse joined up with the trailing edge of the previous one. Each pulse-train solution belonging to the family is characterised by the parameter $W$, the value of the recovery variable at which the leading edge transition takes place (correspondingly, the trailing edge transition takes place at the value ($1-2a-W$) of the recovery variable).
\vskip .5cm
\noindent A periodic pulse-train is equivalent to a pulse propagating along a circular lattice, and can thus be used to model {\it re-entrant} pulses in a ring of excitable cells. Re-entrant waves have recently been the focus of intense attention in understanding the origin of arrythmias and fibrillations in the cardiac tissue, and are thought to constitute the mechanism underlying the break-up of spiral and scroll waves~\cite{new1,new2,new3,new4}.
\vskip .5cm
\noindent
In this context, we shall present, in a future communication, a stability analysis for the pulse solution following an approach adopted in~\cite{ref8}. There we consider the pulse to be travelling on a (sufficiently large) circular lattice (a pulse on a circular lattice is equivalent to a periodic pulse-train on a linear lattice; recall that a pulse-train is characterised by a parameter $W$) and reduce the problem of stability to that of obtaining the spectrum of a mapping in a function space. In deriving the stability criterion, we use a continuum version of \eqref{eq:onea}, \eqref{eq:oneb}, while taking the lattice discreteness into consideration through the relation defining the speed $\chi$ of the pulse in terms of the parameter $W$, a special feature of this relation being the pinning of the pulse at $W=\tilde{a}-a$. We shall examine as to what extent the lattice discreteness with its attendant feature of pinning leads to a tendency of deviation from the so-called restitution hypothesis, the latter being a criterion of stability expressed in terms of the `diastolic interval' (DI) and the `action potential duration' (see, e.g.,~\cite{ref7,ref8,ref9,ref10}). In a separate communication we shall examine certain interesting features of pulse-train propagation on a discrete lattice in the presence of {\it pacemaker} oscillators, wherein the role of the pinning transition will be found to be especially crucial.
\vskip .5cm
\noindent{\bf Acknowledgement:}
One of us (PM) acknowledges with thanks a  research grant from CSIR, India, bearing sanction no. 8/463/Renewal/2004-EMR-I.

\end{document}